\documentclass[11pt]{article}

\usepackage{amsmath,amssymb}
\usepackage{slashed}
\usepackage{xcolor} % Define colors
\usepackage{graphicx}
\usepackage{dcolumn} % Align table columns on decimal point
\usepackage{bm} % Bold math
\usepackage{epsfig}
\usepackage{epstopdf}
\usepackage{grffile}
\usepackage{color}
\usepackage{colordvi}
\usepackage{amsmath,amssymb}
\usepackage{rotating}
\usepackage{lscape}
\usepackage{cite}
\usepackage{float}
\usepackage{hyperref}

\def\Re{{\cal R \mskip-4mu \lower.1ex \hbox{\it e}\,}}
\def\Im{{\cal I \mskip-5mu \lower.1ex \hbox{\it m}\,}}

\def\tev{\,{\ifmmode\mathrm {TeV}\else TeV\fi}}
\def\gev{\,{\ifmmode\mathrm {GeV}\else GeV\fi}}
\def\mev{\,{\ifmmode\mathrm {MeV}\else MeV\fi}}

\newcommand{\inab}{\,{\rm ab}^{-1}}
\newcommand{\infb}{\,{\rm fb}^{-1}}

\newcommand{\gsim}{\raisebox{-0.13cm}{~\shortstack{$>$ \\[-0.07cm] $\sim$}}~}

\newcommand{\beq}{\begin{eqnarray}}
\newcommand{\eeq}{\end{eqnarray}}
\newcommand{\slh}{\hspace{-.5em}/}

\renewcommand{\Re}{{\cal R}}

\newlength{\myVSpace}% the height of the box
\begin{document}
\bibliographystyle{unsrt}
\begin{titlepage}

\begin{center}
{\it }

\end{center}
\vskip 8pt

\begin{center}

\vspace*{15mm}
\vspace{1cm}
{\Large \bf Loop induced singlet scalar production through the vector like top quark  at future lepton colliders}
%Loop induced singlet scalar production through the vector like top quark  at future lepton colliders
\vspace{1cm}

{\small \bf Daruosh Haji Raissi\footnotemark[1], Seddigheh Tizchang\footnotemark[2], and Mojtaba Mohammadi Najafabadi\footnotemark[2] }

 \vspace*{0.5cm}

$\footnotemark[1]$ {\it \small Faculty of Sciences, Department of Physics, Ayatollah Amoli Branch, Islamic Azad University,
Amol,  Mazandaran,  Iran}\\
$\footnotemark[2]$ {\it \small School of Particles and Accelerators, Institute for Research in Fundamental Sciences (IPM) P.O. Box 19395-5531, Tehran, Iran }\\

\vspace*{.2cm}
\end{center}

\vspace*{10mm}

%%%%%%%%%%%%%%%%%%%%             abstract        %%%%%%%%%%%%%%%%%%%%%%%%%%%%    
\begin{abstract}\label{abstract}
%%%%%%%%%%%%%%%%%%%%%%%%%%%%%%%%%%%%%%%%%%%%%%%%%%%%%%%%%%
%
In this paper, we explore the signature of a simplified model
which includes a new singlet scalar state and a vector like quark
at future lepton colliders. In particular, we study the production of 
the new singlet scalar in association with a photon, which proceeds through 
loop level diagrams involving vector like top quark partner, at future $e^-e^+$ colliders with different
center-of-mass energies from 500 GeV to 3 TeV.  To show the sensitivity of the process, 
the exclusion limits on the parameter space of the model are presented considering the decay of the singlet scalar into 
a pair of Higgs boson, followed by the decay of  Higgs bosons into $b\bar{b}$ pairs.
The results are compared to those obtained from the LHC, electroweak precision data and other channels at lepton colliders 
and it is shown that a notable sensitivity to the parameter space of the simplified model
could be achieved.
\end{abstract}
\vspace*{3mm}

%{\bf Keywords}: Beyond Standard Model, Lepton Collider, Vector Like Quark.
\end{titlepage}

\newpage

%%%%%%%%%%%%%%%%%%%%%%%%%%%%%%%%%%%%%%%%%%%%%%%%%%%%%%%                     Introduction          %%%%%%%%%%%%%%%%%%%%%%%%%%%%%%%%%%%%%%%%%%%%%%%%%%
\section{Introduction}\label{sec:intro}
%%%%%%%%%%%%%%%%%%%%%%%%%%%%%%%%%%%%%%%%%%%%%%%%%%%%%%%%%%%%%%%%%%%%%%%%%%%%%%%%%%%%%%%%%%%%%%%%%%%%%%%%%%%%%%%%%%%%%%%%
%
With the Higgs boson discovery at Run I of the LHC \cite{{ATLAS},{CMS}}, followed by
measurements of its properties from the LHC second run  \cite{{massH},{spin-parity},{couplingstrengths},{lifetime}}, 
the spontaneous symmetry breaking mechanism of the Standard Model (SM)  is confirmed.
However, there are still unanswered questions in the SM framework which motivates
proposing theoretical models beyond the SM.
Among them, there are models with extended scalar and fermion sectors of the SM, 
which are well motivated to control the instability of electroweak vacuum \cite{Cao, Xiao:2014kba , Batell} 
and hierarchy problem \cite{Abbas,S.Fajfer}. 
In particular, models of extended quark sector with vector like quarks (VLQs)  appear in 
composite Higgs models \cite{{CHM},{ N. G. Ortiz}}, 
extra dimensions \cite{extraD}, little Higgs models \cite{LittleHiggs}, gauging of the flavor group \cite{Gf}, and
 non-minimal supersymmetric SM \cite{NMSSM}. 
In this work, the concentration is on a simplified model with minimal fields content and interactions, which adds
a singlet scalar and a vector like top quark to the SM. In the considered scenario, both the new scalar and vector like 
top quark mix with the Higgs boson and top quark, respectively \cite{Xiao:2014kba, Dolan:2016eki}.

In the SM context,  based on the next-to-next-to-leading calculation, the Higgs boson
self coupling $\lambda$ tends to be negative at a high energy scale 
of around $10^{10}$ GeV which causes instability of the Higgs vacuum. 
The vacuum stability is quite dependent on how precise the top quark mass is measured.
It is notable that the current measured values of the Higgs boson and top quark masses
suggest that the vacuum could be metastable \cite{vs1,vs2,vs4,vs5,vs6}.
In the considered simplified model, the presence of the new scalar can have a positive contribution to 
the Higgs boson quartic coupling $\lambda$ and consequently could push the Higgs boson 
potential toward the stable phase. It is worth mentioning that the added VLQ in this minimal model
will help preserve the perturbativity of the new scalar quartic coupling \cite{Xiao:2014kba}.

There are tight bounds on the couplings of the light quarks to the heavy vector like quarks from flavor physics \cite{Isidori:2010kg}, 
as a result in this simplified model no interaction among the vector like quark and 
the first two quark generations is considered. The new scalar state couples to the massive 
SM particles through its mixing with Higgs boson and couples to 
the massless photon and gluon via loops
involving the VLQ and top quark
 \cite{{Xiao:2014kba},{Dolan:2016eki},{ H. Han},{ A. Falkowski},{ S. Knapen},{J. Ellis},{ R. Benbrik},{ J. Zhang },{ R. S. Gupta},{ N. Craig}}.

So far, no evidence for existing such 
new particles have been observed by the collider experiments. 
There are studies for vector like top quark $T$ at the LHC by the ATLAS and CMS experiments,
where the searches have been performed for vector like top quark pair $T\bar{T}$ production 
through strong interactions \cite{cms-VLQ,atlas-VLQ,atlas-VLQ2}.
In these searches,  the $T$ and $\bar{T}$ quarks are assumed to decay
into $tZ$, $tH$, and $Wb$ and the searches have been performed in different channels like single lepton,  
multilepton,  and full hadronic.
In all studied final states, the results are
compatible with the SM expectations hence limits have been set on the model parameters.
Other studies on constraining the masses, couplings of 
the vector like top quark $T$, and the new scalar
can be found in Refs.\cite{Mathieu, Giacomo, Giacomo2, Saavedra, Saavedra2, Dolan:2016eki, vl1,vl2}.
Model-independent approach using effective operators for the vector like top quark partner has been studied in Ref.\cite{eff}.
The vector like top quark $T$ can also be produced singly in association with a light quark and either a top quark or a bottom quark at the
LHC which has been the subject of several studies such as Refs.\cite{Sirunyan:2017ynj ,Aaboud:2018saj}.

In addition to the direct searches, the parameter space of the simplified model with a singlet scalar and VLQ has been probed
using the electroweak precision observables as well as the Higgs coupling precision measurements 
that could be found in Ref.\cite{Xiao:2014kba}. The requirement of vacuum stability and the unitarity of the 
VLQ and scalar scatterings constrain the parameter space. Among all the limits, those from vacuum stability are the tightest \cite{Xiao:2014kba}.

The future electron-positron colliders such as International Linear Collider (ILC) \cite{Fujii:2015jha, Barklow:2015tja},
Compact Linear Collider (CLIC) with a possibility of going to high 
center-of-mass energy up to 3 TeV \cite{CLIC:2016zwp}, the Future Circular Collider FCC-ee \cite{Gomez-Ceballos:2013zzn,Koratzinos:2015hya},
  and Circular Electron-Positron Collider CEPC \cite{CEPC-SPPCStudyGroup:2015csa}
provide  very clean places to measure the SM parameters and to search for new physics effects. 
Lack of hadronic initial state,  low amount of background, and accurate knowledge of initial beam energy
would flatten the way of precision studies or discovery of new particles.

The goal of the present work is to study the phenomenology of 
the direct production of the singlet scalar at the future lepton colliders and inquire its effects on the phenomenology 
of lepton colliders. In particular, the production of the singlet scalar in association with a photon
is studied because the final state  has an energetic photon and
could be used as a handle to reduce the background contributions and trigger the signal events. 
The process proceeds through loop level diagrams where the VLQ contributes to the loops, therefore the channel 
is sensitive to the related parameters of the VLQ and would be a complementary process to the other
channels to search for the model.
This paper is organized as the following:
In section \ref{sec:model} a brief review of the model and its main properties are given.
Section \ref{sec:scalarproduction} provides a phenomenological study on the new scalar associated with a photon
at the high energy lepton colliders. In section \ref{sec:constrains}, the possible final states and potential bounds
on the model parameters are presented. The results are compared with the constraints obtained 
from the LHC data and from the expectation of the single scalar production channel at a future high energy lepton collider. 
 Section \ref{sec:conclu} is devoted to the summary and conclusions.

%%%%%%%%%%%%%%%%%%%%%%%%%%%%%%%%%%%%%%%%%%%%       The model        %%%%%%%%%%%%%%%%%%%%%%%
\section{The model} \label{sec:model}
%%%%%%%%%%%%%%%%%%%%%%%%%%%%%%%%%%%%%%%%%%%%%%%%%%%%%%%%%%%%%%%%%%%%%%%%%%%%%
%
 In this section, a brief description of the simplified model where the SM is extended by
 adding a new neutral singlet scalar $S$ and a vector like quark $T$  is given.
 The VLQ $T$ carries the same quantum numbers as the  right-handed top quark 
 and mixes only with the SM top quark.
More detailed description of the model could be found in Refs. \cite{{Xiao:2014kba},{Dolan:2016eki}}.  
In the new scenario, the scalar, Yukawa and gauge sectors of the SM Lagrangian receive changes:
\begin{equation}
\mathcal{L}\supset\mathcal{L}_{\rm scalar}+ \mathcal{L}_{\rm Yukawa}+ \mathcal{L}_{\rm gauge},
\end{equation}
where
\begin{eqnarray}\label{eq:potential}
\mathcal{L}_{\rm  scalar}&=& \frac{1}{2}( D_{\mu}H)^{\dagger}( D^{\mu}H)+ \frac{1}{2} \partial_{\mu}S ~\partial^{\mu}S-\mu^2 H^\dagger H + \lambda (H^\dagger H)^2 \\\nonumber
&+&  \frac{a_1}{2}H^\dagger H \, S + \frac{a_2}{2} H^\dagger H \, S^2\ + b_1 S + \frac{b_2}{2} S^2 + \frac{b_3}{3} S^3 + \frac{b_4}{4} S^4,
\end{eqnarray}
here $H$ is the SM Higgs boson doublet and the new scalar 
field is denoted by $S$.
To keep the Yukawa term $S\bar T T$, no $Z_2$ symmetry is applied however conventionally $ a_{1}, b_{1},b_{3}$ 
are set to zero. With such assumptions on the couplings, one can explain all the relevant measurements and also explain the
motivations for which the model has been proposed. 
Both the SM Higgs doublet and new scalar $S$ acquire non-zero vacuum expectation values:
\begin{equation}\label{eq:vev}
H = 
\begin{pmatrix}
i \phi^+ \\
\frac{1}{\sqrt2} ( v_H + h  + i\phi^0)
\end{pmatrix},\,\,\,\,\, \quad S = (s + v_S),
\end{equation}
where $v_H$ is the vacuum expectation value ({\it vev}) of the SM Higgs boson and is $\approx 246~{\rm GeV}$ and 
the {\it vev} of the singlet scalar is denoted by $v_{S}$.
After the spontaneous symmetry breaking and expanding the Lagrangian around its minimum, the squared mass matrix has the following form:
\begin{equation}
\mathcal{M}^2_{\rm scalar}=
\begin{pmatrix}
2 \lambda v_H^2 & a_2\, v_H v_S \\
a_2 v_H v_S &2\,b_4 v_S^2
\end{pmatrix},
\end{equation}
and the squared masses of the physical eigenstates are found to be:
\begin{eqnarray}
m^2_{h_1,h_2}=\lambda\, v_H^2+b_4\,v_S^2\mp \sqrt{(b_4\,v_S^2-\lambda v_H^2)^2+a_2^2\,v_H^2 v_S^2}.
\end{eqnarray}
The physical eigenstates $h_{1,2}$ are related to the singlet scalar field  $s$ and  the SM Higgs field $h$ through the
following transformation:
\begin{eqnarray}
\begin{pmatrix}
h_1 \\ h_2
\end{pmatrix}
=\begin{pmatrix}
 \cos(\theta) & - \sin(\theta) \\ \sin(\theta) &\cos(\theta)
\end{pmatrix}
\begin{pmatrix}
h \\s
\end{pmatrix},
\end{eqnarray}
where $\theta$ is the mixing angle and is defined as:
\begin{eqnarray}
\tan (2\theta)=\frac{a_2\, v_H v_S}{b_4 v_S^2-\lambda\, v_H^2}.
\end{eqnarray}
It is assumed that the $h_{1}$ scalar field is  the SM Higgs boson with a mass of  $m_{h_1} = 125$ GeV, which will denoted as $h$ afterhere.
To have the vacuum stability at large energy scales, the mixing angle $\theta$ has to be small\cite{{Ilnicka},{Grace}}.
Considering the LHC measurements of the Higgs boson, electroweak precision data, and respecting the vacuum
stability conditions impose that  $|\sin(\theta)| \lesssim 0.2$.

The third quark generation Yukawa part of the Lagrangian is modified as follows: 
\begin{eqnarray}\label{eq:lagrang Yukawa}
\mathcal{L}_{\rm Yukawa} &=& y_T S\overline{T}^{\rm int}_L T^{\rm int}_R + y_t \overline{Q}^{\rm int}_L\widetilde{H} t^{\rm int}_R + y_b \overline{Q}^{\rm int}_L H b_R + \lambda_T \overline{Q}^{\rm int}_L \widetilde{H} T^{\rm int}_R,  \nonumber
\end{eqnarray}
where int index stands for weak interaction eigenstates, 
$\bar{Q}_L$  is  left-handed third generation quark doublet, $\bar{Q}_L= (\bar{t}^{\rm int}_L \; \bar{b}^{\rm int}_L)$ 
and  $\widetilde{H} = i\sigma_2 H^*$. 
We note that one could add a Dirac mass term for vector like top quark $T$ to the above Lagrangian however
after the spontaneous symmetry breaking  $T$ acquires mass.
Therefore, in order to reduce the number of free parameters in the model, Dirac mass term is not added to the Lagrangian. 
A term proportional to $ \overline{T}^{\rm int}_L t^{\rm int}_R$, can also be present in the $\mathcal{L}_{\rm Yukawa}$, however,
it can be removed by a redefinition of the right-handed fields $(t^{\rm int}_R, T^{\rm int}_R)$ \cite{Dawson:2012di}. 
After the spontaneous symmetry breaking, the SM top quark $t^{\rm int}$ and vector like top quark $T^{\rm int}$ 
mix and the mass matrix can be written as:
\begin{equation}
\mathcal{M}_{\rm Yukawa}=
\begin{pmatrix}
y_tv_H/\sqrt{2} & \lambda_Tv_H/\sqrt{2} \\
0 &y_T v_S
\end{pmatrix}.
\end{equation}
The mass matrix is diagonalized by the unitary transformation with $\theta_L$ and $\theta_R$ rotation angles:
\begin{equation}\label{eq: bi-unitary}
\begin{pmatrix}
t_{L/R}\\ T_{L/R}
\end{pmatrix} 
= \mathbf{U}_{L/R}
\begin{pmatrix}
t^{\rm int}_{L/R}\\ T^{\rm int}_{L/R}
\end{pmatrix},
\end{equation} 
where $t_{L/R}$ and $T_{L/R}$ are the physical mass eigenstates and the unitary matrices are written as:
\begin{equation}
\mathbf{U}_{L/R}  = 
\begin{pmatrix}
\cos\theta_{L/R} & -\sin \theta _{L/R} \\
\sin \theta_{L/R} & \cos\theta_{L/R}
\end{pmatrix} .
\end{equation}
The squared mass eigenvalues for the SM top quark and the vector like quark $T$ are found to be:
\begin{equation}\label{mass eigenstate}
m_{T,t}^2  =   \frac14\left( y_t^2 v_H^2 + \lambda_T^2 v_H^2 + 2 y_T^2 v_S^2  \pm \sqrt{(y_t^2 v_H^2 + \lambda_T^2 v_H^2 + 2 y_T^2 v_S^2)^2-8 y_t^2 v_H^2 y_T^2 v_S^2}\right), 
\end{equation}
where the lighter eigenstate is assumed to be the SM top quark with $m_{t}=173.2\,{\rm GeV}$ 
and the heavier is taken as the vector like $T$ with a mass $m_T$.
Two mixing angles are related through the following relation \cite{Aguilar}:
\begin{equation}\label{mixing angels}
\tan (\theta_R)=\frac{m_t}{m_T}\tan{(\theta_L)}.
\end{equation}
It is clear that the left-handed mixing angle is always dominant, in particular for the heavy vector like $T$.
The  Yukawa coupling terms, mass terms of the $t,T$ and the mixing 
term between $t$ and $T$ are given by:
\begin{eqnarray}\label{eq:yukawa-lagrang2}
\mathcal{L}_{\rm Yukawa}  &\supset & \frac{m_T}{v_Hv_S}\left(s_L^2v_S\,(h-i\phi^0) + c_L^2v_H\,s\right)\overline{T}_L T_R \\ \nonumber
&+ & \frac{m_t}{v_H v_S}\left(c_L^2v_S\,(h-i\phi^0) + s_L^2v_H\,s\right)\bar{t}_L t_R \\ \nonumber 
& + & \frac{m_T}{v_Hv_S}s_Lc_L\left(v_S\,(h-i\phi^0) - v_H\,s\right)\bar{t}_L T_R\\ \nonumber
& + & \frac{m_t}{v_Hv_S}s_Lc_L\left(v_S\,(h-i\phi^0) - v_H\,s\right)\overline{T}_L t_R, \nonumber 
\end{eqnarray}
where $s_L(c_L) \equiv \sin\theta_L (\cos\theta_L)$.
Electroweak gauge interactions of vector like quark $T$ with the quantum number $Q_T=2/3$ and $Y_T=4/3$ with the SM third quark generation $t, b$ 
are as follows:
\begin{eqnarray}
 \mathcal{L}_{\rm gauge} & \supset&  i\,\bar{t}\slashed{\partial}t  + i\,\bar{b}\slashed{\partial} b +  i\,\overline{T}\slashed{\partial} T  \\\nonumber 
&+ & e \left (Q_{t} \bar{t} \gamma^\mu t + Q_{T} \overline{T} \gamma^\mu T + Q_{b} \bar{b} \gamma^\mu b \right)A_\mu\\\nonumber
&+&  \frac{g}{\sqrt{2}}\left (( c_L\bar{t} \gamma^\mu P_L b + s_L\overline{T} \gamma^\mu P_L b)W_\mu^+ +  (c_L \bar{b}\gamma^\mu P_L t + s_L \bar{b}  \gamma^\mu P_L T)W^-_\mu \right )\\ \nonumber
&+ & \frac{g}{c_w}\left(  \overline{T}\gamma_\mu \left (\frac{s_L^2}{2} P_L - Q_{T}s_w^2\right) T  + \bar{t} \gamma_\mu \left (\frac{c_L^2}{2} P_L - Q_{t}s_w^2\right) t  \right. \\\nonumber 
& +& \left. \bar{b} \gamma_\mu \left (-\frac12 P_L - Q_{b}s_w^2\right)b  + \bar{t} \gamma_\mu \frac{s_Lc_L}{2} P_L  T + \overline{T} \gamma_\mu \frac{s_Lc_L}{2} P_L  t \right) Z_\mu, \nonumber
\end{eqnarray}
where $\theta_w$ is Weinberg weak mixing angle, $s_w(c_w) \equiv \sin\theta_w (\cos\theta_w)$ and $P_L=(1-\gamma_5)/2$ is projection operator.  
More explanation for driving of the above interactions are given in appendix of Ref.\cite{Dolan:2016eki}.
The simplified model followed here has five unknown parameters which consists of 
the mass of  vector like top quark $m_{T}$,  the new singlet scalar mass $m_{h_2}$,
vacuum expectation value of the singlet scalar $v_{S}$, and the mixing angles $\theta_{L}$ and
$\theta$ in the fermion and scalar sectors, respectively.

In Ref.\cite{Buttazzo:2018qqp},  the production of a new scalar singlet $h_{2}$  has been studied via vector boson fusion ($e^{-}+e^{+} \rightarrow h_{2}+\nu\bar{\nu}$)
 at a high energy lepton collider with the center-of-mass energy of 3 TeV. 
 In particular, it has been found that a future high energy lepton collider would be able to examine
 the single scalar production rate with a few tens of atto-barn. 
The production of the scalar $h_{2}$ associated with a $Z$ boson at a lepton collider operating 
at the center-of-mass energy of 240 GeV has been studied in Ref.\cite{Chang:2017ynj}.
In section \ref{sec:scalarproduction},  we calculate the production cross section of the
scalar $h_{2}$ in association with a photon at the lepton colliders. The cross section
is presented at different center-of-mass energies of the electron-positron collisions and its dependence on the free parameters of the model
is presented.

%

%%%%%%%%%%%%%%%%%%%%%%%%%%%%%%%%%%%%%%%%%%%%%%%%%       The associated production of singlet scalar with photon        %%%%%%%%%%%%%%%%%%%%%%%%%%%%%%%%%%%%%%%%%%%%
\section{Singlet scalar production in association with a photon}\label{sec:scalarproduction}
%%%%%%%%%%%%%%%%%%%%%%%%%%%%%%%%%%%%%%%%%%%%%%%%%%%%%%%%%%%%%%%%%%%%%%%%%%%%%%%%%%%%%%%%%%%%%%%%%%%%%%%%%%%%%%%%%%%%%%%%

In this section, we propose an alternative way to have access to parameter space of the
simplified model by  considering the process $e^{-}e^{+} \rightarrow h_{2}\gamma$ which proceeds via loops
with contributions from SM fermions, gauge bosons and, the VLQ. 
In Ref.\cite{abdolhak}, the cross section of the associated production of a photon and a Higgs boson in the context of 
Minimal Supersymmetric Standard Model (MSSM) has been calculated. The production rates for 
the associated production of both the CP-even and CP-odd Higgs bosons of the MSSM have been studied.
 For the CP-even MSSM Higgs
boson production with a photon, other s-channel Feynman diagrams involving loops with
charginos, charged Higgs bosons, squarks, and sleptons  appear and in the t-channel box diagrams
chargino/sneutrino and neutralino/selectron contribute to the production process.  
The production rate of the Higgs boson in association with a photon at electron-positron colliders,
in the context of extended Higgs models, like the two-Higgs-doublet model, the inert doublet model,  and the inert triplet model (ITM) has
been studied in Ref.\cite{Kanemura:2018esc}.  The authors found that the charged scalars of these models via loop diagrams can
generate sizable contributions to the production cross section of $h+\gamma$.  
The potential of the LHC to probe the new physics effects in the SMEFT (SM Effective Field Theory) framework through 
the Higgs boson production associated with a photon has been studied in Ref.\cite{kkm}.
The next-to-leading order QCD corrections to the production of a Higgs boson associated with a photon
has been calculated in Ref.\cite{Sang:2017vph}, where it has been shown that these corrections could increase the production 
rate up to $20\%$.

\begin{figure}
           \center
	\includegraphics[width=0.98 \linewidth]{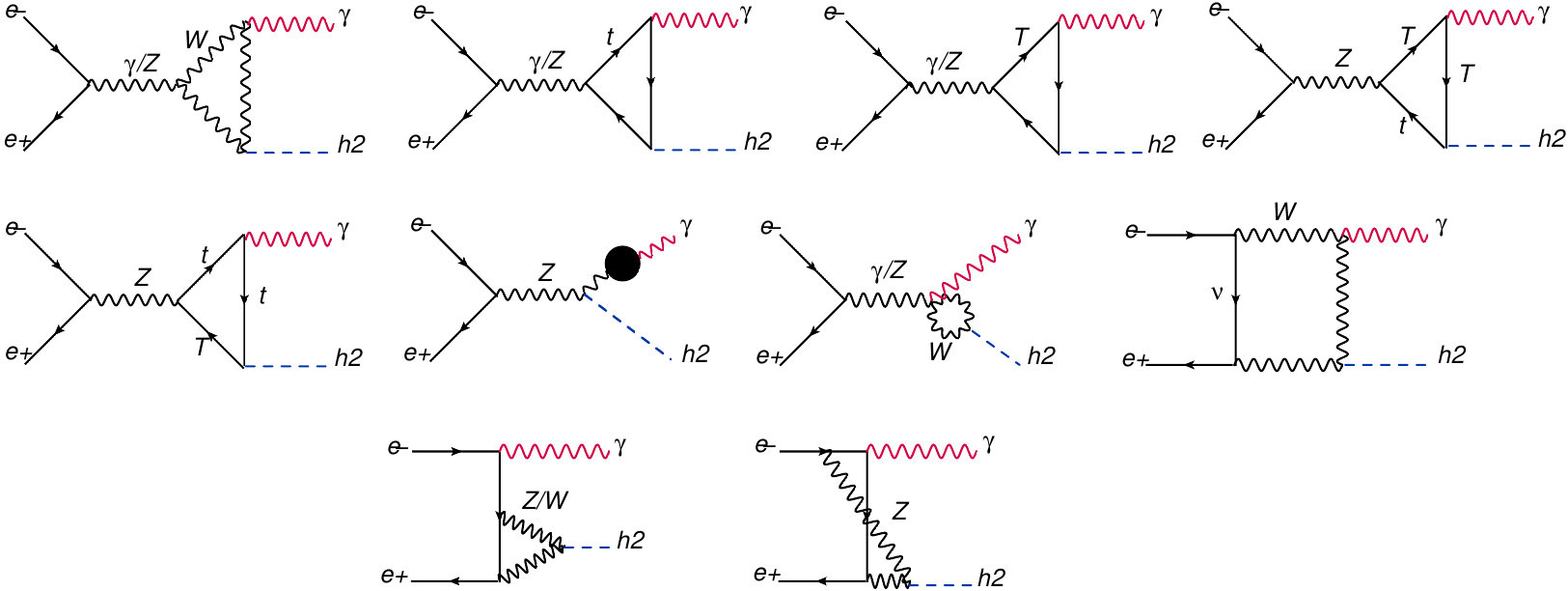}
		\caption{\small  Representative Feynman diagrams for the process $e^{-}e^{+} \rightarrow h_{2}  \gamma$ in the context of simplified model. }
	\label{feyn}
\end{figure}

Within the simplified model, the $e^{-}e^{+} \rightarrow h_{2}\gamma$ process proceeds through loop-level Feynman diagrams which include  SM fermions and gauge bosons
as well as the additional contributions from the vector like top quark.
Representative Feynman diagrams contributing to the process $e^-e^+\rightarrow h_2 \gamma$ are presented in Fig.\ref{feyn}. 
There are s-channel diagrams with $Z/\gamma$ boson exchange 
in which virtual $W$ boson, heavy SM fermion  (mostly top and bottom quarks) as well as the
vector like top quark in the loops are involved. 
The new singlet scalar $h_{2}$ couples to the top quark via both the fermion and scalar mixing which 
causes triangle diagrams in the s-channel production.
The t-channel Feynman diagrams involve $W,Z,\nu_{e}$ and electron exchanges. 
Contributions from s-channel diagrams with the SM Higgs boson $h$ and its interference with the
$Z$ boson or photon is negligible.

Since the process occurs at higher order electroweak interaction, the cross section is expected to be rather small, however
the signal is very clean specially due to the presence of an energetic photon in the final state. This would allow
to achieve a reasonable background suppression and leads to have a good sensitivity in particular using the expected 
large amount of data by the future lepton colliders.

Within the considered simplified model in this work, the one-loop  amplitude, neglecting the mass of electron,  for the $h_{2}+\gamma$ production 
can be written as the sum of the amplitudes of all contributing diagrams:
\begin{equation}
{\cal M}=\sum_{k=1,2}\sum_{v=+,-} \Lambda_k^v C_k^v,
\label{fullamplitude}
\end{equation}
where $\Lambda_k^{\pm} $ have the following form \cite{abdolhak}:
\begin{eqnarray}
\Lambda^{\pm}_{1} &=& \overline{v}(q_+)(1 \pm \gamma_5)
\big( \epsilon\slh^*_\gamma \; p_{\gamma}.q_{-}
       - p\slh_{\gamma}  \epsilon_\gamma^*.q_{-} \big) \;u(q_-) \nonumber\\
\Lambda^{\pm}_{2} &=& \overline{v}(q_+)\; (1 \pm \gamma_5)
 \big( \epsilon\slh^*_\gamma \; p_{\gamma}.q_{+}
       - p\slh_{\gamma} \; \epsilon_\gamma^*.q_{+} \big) \;u(q_-),
\label{amplitude}
\end{eqnarray}
where $q_{\pm}$ are the momenta of the $e^\pm$ beams, $p_\gamma$ is the  four-momentum of 
external photon and $\epsilon_\gamma$ is the corresponding polarization vector and $C_{k}^{\pm}$ represent  
form factor coefficients obtained by summing the diagrams depicted in Fig.\ref{feyn}:
\begin{equation}
  C_{k}^{\pm} = C^{\gamma\pm}_k +  C^{Z\pm}_k + C^{\pm _{\rm Wbox}}_{k}+ C^{\pm_{\rm Zbox}}_{k},
   \label{formfactor} 
\end{equation}
where $ C^{\gamma, \pm}_k $ and $C^{Z,\pm}_k$ represent the contributions of  $\gamma$ and $Z$ propagators in s-channel diagrams for vertex corrections and $C^{\pm _{\rm Wbox}}_{k}$ and $ C^{\pm_{\rm Zbox}}_{k}$ are the contributions of box diagrams.
The total cross section for unpolarized beam is obtained as:
\beq
\frac{{\rm d}\sigma}{{\rm d}\cos\theta_s} = 
\frac{s-m_{h_2}^2}{64 \pi s} \frac{1}{(16\pi^2)^2}   
\bigg[ u^2\; (|C_1^+|^2+ |C_1^-|^2 ) + t^2\; (|C_2^+|^2+|C_2^-|^2)\bigg],
\eeq
where $\theta_s$ is the the angle between the incoming electron and outgoing photon.
To obtain the above differential cross section, an averaging over the helicities of the incoming leptons and a sum over the
outgoing photon polarization has been performed. $s,t$ and $u$ are the Mandelstam variables and are defined as
 $s=(q_{-}+q_{+})^2$, $t=(q_{+}-p_\gamma)^2$ and $u=(q_{-}-p_\gamma)^2$.
 The $C_k^{\pm}$  coefficients can be written as a function of Passarino-Veltman  functions.
 The $C^{\gamma\pm}_k$ coefficient contains contribution from three diagrams involving $W$ boson, top quark and vector like top quark 
 in the loop-induced vertex $h_{2}\gamma\gamma $ and the $C^{Z\pm}_k$ includes $W$ boson, top quark and vector like top quark contribution
 in loop-induced vertex $h_{2}Z \gamma$.  $C_k^{\gamma \pm}$ and  $C^{Z\pm}_k$ have the following forms:
\begin{eqnarray}
C^{\gamma\pm}_1 &=& C^{\gamma\pm}_2 =
- \frac{e}{2}\,\frac{1}{s}\; G^{\gamma}, \nonumber\\ 
C^{Z\pm}_1 &=& C^{Z\pm}_2 =
-\frac{e\, z^{\pm}}{4 s_w c_w}\, \frac{1}{s-M_Z^2}\; G^{Z}, 
\end{eqnarray}
where $e$ is the size of electron electric charge, $z^{+} = -1+2s^{2}_{w}$, and $z^{-} = 2s^{2}_{w}$. Considering both the bosonic and fermionic contributions in s-channel vertex correction, $G^\gamma$ and $G^Z$ are obtained as follows:
\begin{eqnarray}\label{formfi}
G^\gamma &=& \frac{e^3 M_W}{ s_w}\Bigg[F^{\gamma, W} s_\theta-\sum_{f}\, 4\,Q_f^2\,N_c\,\frac{m_f^2}{M_W^2} \,F^f s_\theta  \nonumber\\
&- &4\, F^t \frac{ m_t^2}{M_W^2} N_c ~ Q_t^2  \left(r\,  s_L^2\, c_\theta +c_L^2\,  s_\theta \right)
- 4\, F^T \frac{ m_T^2}{M_W^2} N_c ~ Q_T^2 \left(r\,  c_L^2\, c_\theta + s_L^2\,s_\theta\right)\Bigg],\nonumber\\
G^{Z} & = & \frac{e^3\,M_W}{c_w\,s_w^2} \Bigg[
\,F^{Z,W} s_\theta + \sum_{f}\, 2~Q_f\,N_c\, \frac{m_f^2}{M_W^2} 
(I^f_3-2~s_w^2 Q_f)  \,F^f s_\theta \nonumber\\
&+& 2~Q_t\,N_c\, \frac{m_t^2}{M_W^2} 
(I^t_3 c_L^2-2\, s_w^2 Q_t)(c_L^2\,s_{\theta}+r\, s_L^2\,c_\theta)  \,F^t 
\nonumber\\
&+&2~Q_T\,N_c\, \frac{m_T^2}{M_W^2}~(I^T_3 s_L^2-2 s_w^2\, Q_T)(s_L^2\, s_{\theta}
+ r\, c_L^2\, c_\theta) \,F^T \nonumber\\
&+&2 \, Q_t\,N_c\, \frac{m_t+m_T}{M_W^2}~s_L^2\, c_L^2\,(s_\theta -r\,  c_\theta) F^{n}\nonumber\\
&+&2 \, Q_T\,N_c ~\frac{m_t+m_T}{M_W^2}~s_L^2\, c_L^2\,(s_\theta -r\,  c_\theta) F^{l}\Bigg],
\label{gizz}
\end{eqnarray}
where $m_f$, $Q_f$ and $I_3^f$ are the mass, electric charge and third component of weak isospin of the 
fermion $f$ ($f$ can be all fermions except for $t$ and $T$), respectively.
$N_c$ is the number of  QCD colors and $r\equiv v_H/v_S$ and $s_{\theta}~(c_{\theta}) \equiv \sin \theta~(\cos \theta)$.  Functions $F$ with various indices in Eq.\ref{gizz} are 
the combination of  Passarino-Veltman  scalar functions and are given in the Appendix \ref{app}.
In order to ensure all the contributing Feynman diagrams from SM and new physics are consistently included, 
the gauge invariance of the matrix elements is checked through validating the Ward-Takahashi identity.
In the model, the new scalar couples to the SM particles through mixing with the Higgs boson as 
 $ h= c_{\theta}\, h_1+s_{\theta}\, h_2$.
Therefore, the new scalar coupling to weak gauge bosons and fermions (except the top quark) are similar to 
the SM Higgs coupling and just receive a correction factor $s_{\theta}$. 
The scalar coupling to top quark is modified by both mixing scalar and Yukawa top sectors (see Eq.\ref{eq:yukawa-lagrang2}).
Among the diagrams, the amplitude of box diagrams ($C^{\pm W,Z\rm{ box}}_k$) have the same form as 
the SM Higgs associated production with photon by only replacing $m_H\rightarrow m_{h_2}$ and
adding $s_{\theta}$ which comes from $VVh_2$ coupling. The explicit forms of $C^{\pm W,Z\rm{ box}}_k$  as a function of Passarino-Veltman scalar functions are given in Appendix\ref{app}.
While in triangle diagrams,  vertices $\gamma \gamma h_2$ and $Z \gamma h_2$ are modified 
due to the contribution of singlet scalar couplings to top quark, $W$ boson, and  the vector like top quark partner.
Comparing the box and triangle diagrams contributions, we find the box diagram contributions are expected to be suppressed  significantly 
at small scalar mixing angle.
Package-X \cite{Hiren} and LoopTools \cite{looptools} are used to reduce the tensor integrals and to evaluate the one loop Feynman integrals.
The Passarino-Veltman formalism according to Ref. \cite{abdolhak} is employed in this work.

\begin{figure}[h!]
 \centering
	\includegraphics[width=0.60 \linewidth]{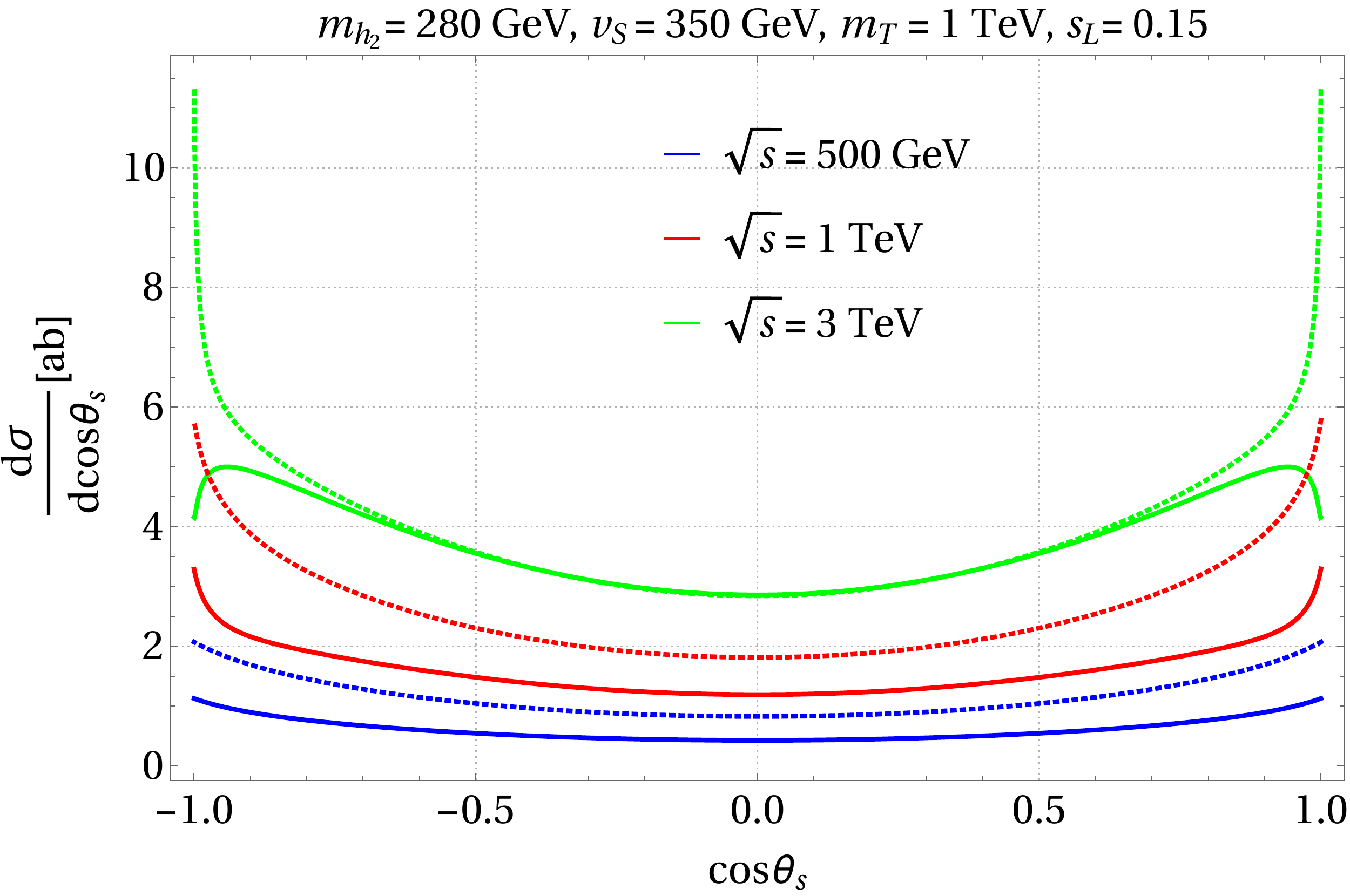}
	\caption{\small  Differential cross section of $h_2+ \gamma$  production  as a function of scattering angle $\cos\theta_s$ for $m_{h_2}=280\,$GeV
	   for different center-of-mass energies with $s_{\theta}=0.15$ (solid) and $s_{\theta}=-0.15$ (dotted).}
	\label{difsigma}
\end{figure}

The differential cross section $d\sigma(e^-e^+\rightarrow  h_2\,\gamma)/d\cos\theta_s$  
 for three center-of-mass energies $\sqrt{s}=500\,$GeV and $1,3\,$TeV are presented in Fig.\ref{difsigma}.
The distributions are shown for two values of $s_\theta = \pm 0.15$ which are denoted by solid ($+0.15$) and dotted ($-0.15$)
curves.
We see that the cross section is not symmetric on $s_\theta$ which was expected as the cross section $e^- e^+ \rightarrow  h_2~\gamma$ 
is  proportional to $\propto(A s_\theta + B c_\theta)^2$. 
Obviously, the cross section is considerably sensitive to the sign of  $s_\theta$ and larger cross section is expected for the
negative values of $s_\theta$ in particular for the low center- of- mass energies of 500 GeV and 1 TeV.
Moreover, the effect of the sign of  $s_\theta$ is
negligible at small scattering angles at $\sqrt s=3\,$TeV.  One can also see that the angular distribution for both signs of $s_\theta$
is symmetric on the scattering angle and no forward-backward asymmetry is expected.

\begin{figure}[h]
	\center
	\includegraphics[width=0.6 \linewidth]{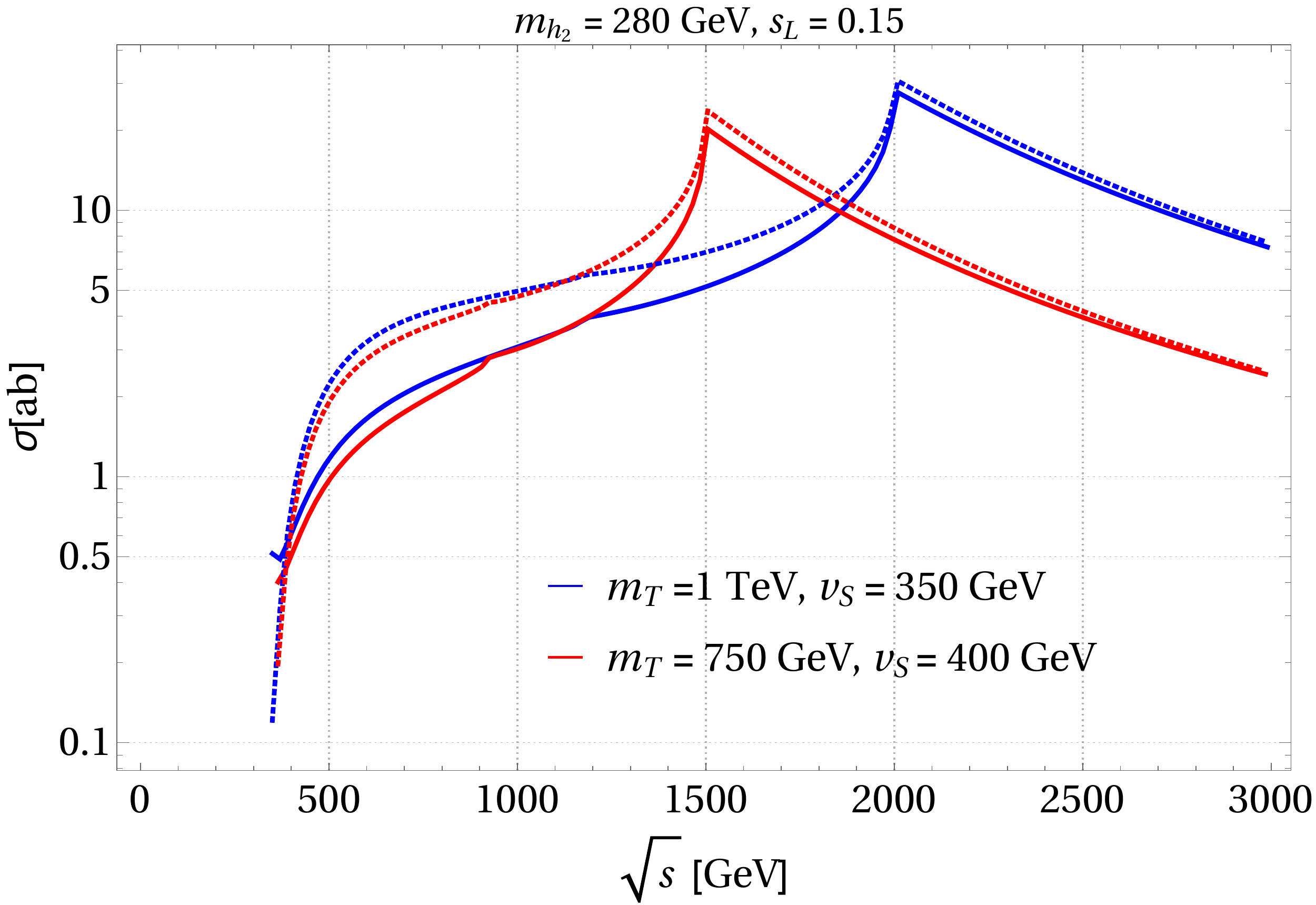}
	\caption{\small Total cross section of  $h_2+ \gamma$ production  as a function of center-of-mass energy $\sqrt s$ for different values of $m_T$ and $v_S$ with  $s_{\theta}=0.15$ (solid) and $s_{\theta}=-0.15$ (dotted).}
	\label{sigma}
\end{figure}

In Fig.\ref{sigma},  the total cross section of $e^-e^+\rightarrow h_2\,\gamma$ 
is presented as a function of center- of-mass energy $\sqrt{s}$ for $m_{h_2} = 280$ GeV, $s_{L} = 0.15$, and $s_{\theta} = \pm 0.15$.
The cross sections are shown for two scenarios of assumption on the two sets of free parameters: $(m_{T}, v_{S}) = (1000, 350), (750, 400)$ GeV.
For both scenarios, the maximum value of cross section occurs at $\sqrt{s} \sim 2 m_{T}$.
As can be seen, the cross section increases rapidly up to $T\bar{T}$ threshold then drops slightly with increasing the center-of-mass energy like 
$1/s$. It is notable that the contribution of s-channel diagrams to the total cross section is dominant with respect to the box and t-channel diagrams. 
The impact of the sign of $s_\theta$ is explicit in particular for the center-of-mass energies less than  $T\bar{T}$ threshold, {\it i.e.} $2m_{T}$. 
While as the center-of-mass energy increases,  the effect of the sign of $s_\theta$ is hardly distinguishable. In addition, 
the cross section enhances when $m_T$ goes up.

\begin{figure}
	\center
	\includegraphics[width=0.6 \linewidth]{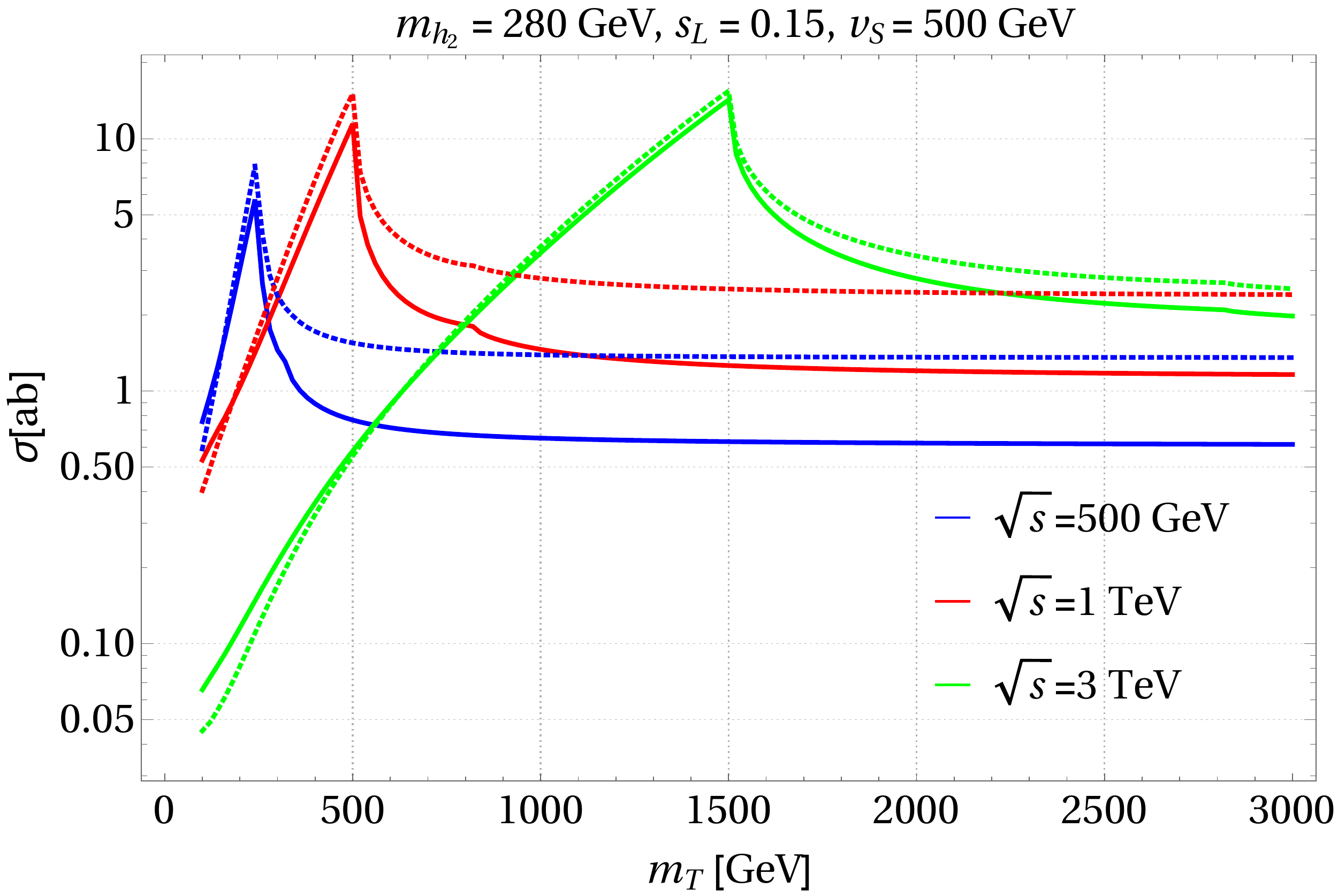}
	\caption{\small Total cross section of $h_2+ \gamma$ production  as a function of vector like
	 top quark mass, $m_T$, for different values of  center-of-mass energies with $s_{\theta}=0.15$ (solid) and $s_{\theta}=-0.15$ (dotted).}
	\label{sigmamT}
\end{figure}

\begin{figure}[h!]
\center
	\includegraphics[width=0.45 \linewidth]{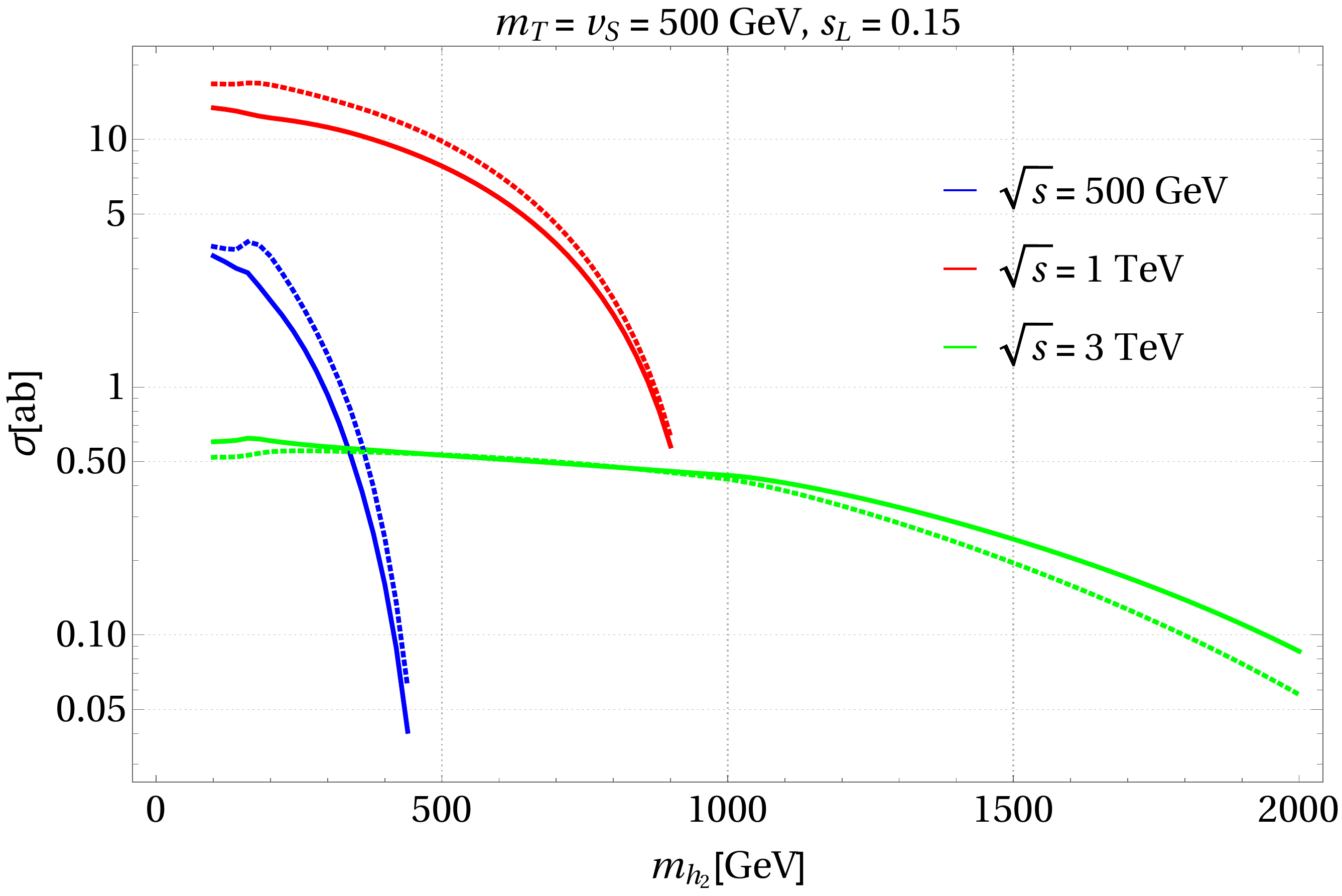}
	\includegraphics[width=0.45 \linewidth]{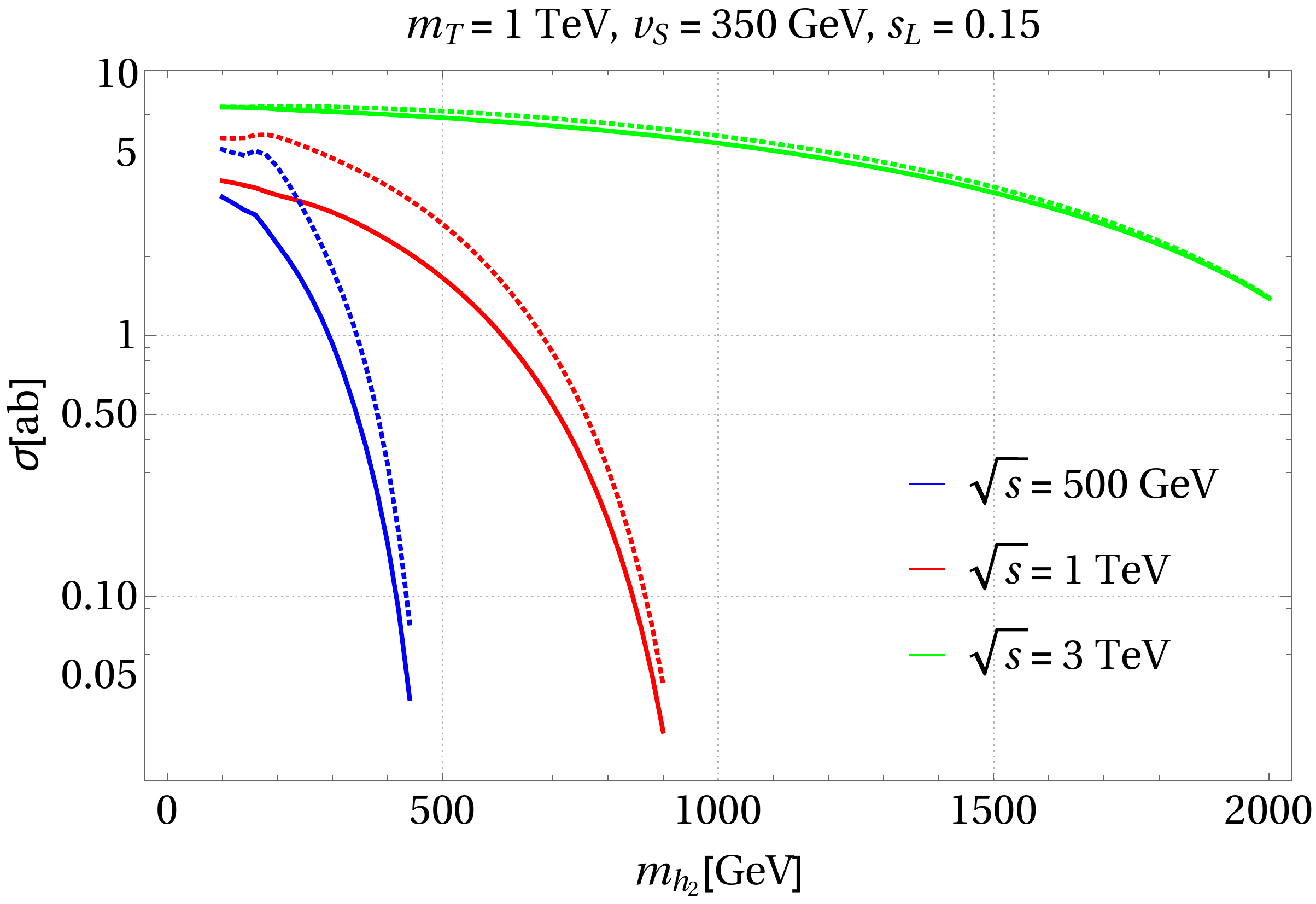}
	\caption{\small  The total cross section of $h_2+ \gamma$  production  as a function of scalar mass $m_{h_2}$ for three center-of-mass energies 
	 with $m_T=v_S=500\,$ GeV  (left)  and $m_T= 1 $ TeV, $v_S=350\,$ GeV (right) and for $s_{\theta}=0.15$ (solid) and $s_{\theta}=-0.15$ (dotted).}
	\label{sigma-mh2}
\end{figure}

To illustrate the dependence of the cross section of  $e^-e^+\rightarrow h_2 \gamma$ 
process on the mass of vector like top quark, we show the cross section as a function of $m_{T}$ in Fig.\ref{sigmamT}. 
The cross section are shown for three center-of-mass energies of 500 GeV, 1 TeV, and 3 TeV for two 
cases of the sign of mixing angle of $s_{\theta}= \pm 0.15$.
The plot shows that the cross section increases quickly with $m_{T}$  up to 
$\sim m_T=\sqrt{s}/2$ then it remains almost constant. The cross section is larger for the
larger center-of-mass energy. 
 Fig.\ref{sigma-mh2} shows the cross section in terms of of the mass of new scalar
  $m_{h_2}$ for $\sqrt{s}=500\,$GeV, $1$ and $3\,$TeV. 
 Because of the larger phase space, the production cross section is large for the
 low mass of scalar and it decreases by increasing $m_{h_2}$. Therefore, more sensitivity is expected 
 to the regions in parameter space with a light scalar. We also note that the cross section 
 decreases more rapidly for the center-of-mass energies of 500 GeV and 1 TeV than the 3 TeV case.

%%%%%%%%%%%%%%%%%%%%%%%%%%%%%%%%%%%%%%%%%%%%%%%%%%      Constrains      %%%%%%%%%%%%%%%%%%%%%%%%%%%%%%%%%%%%%%%%%%%%%%%%%%%%%
\section{Experimental signatures and possible constraints}\label{sec:constrains}
%%%%%%%%%%%%%%%%%%%%%%%%%%%%%%%%%%%%%%%%%%%%%%%%%%%%%%%%%%%%%%%%%%%%%%%%%%%%%%%%%%%%%%%%%%%%%%%%%%%%%%%%%%%%%%%%
%

There are various  direct searches for the new scalar $h_{2}$ at the hadron and lepton colliders. Of particular interest in this work is to perform the search  for $h_{2}$  via 
its production in association with a photon at the future lepton colliders.

\begin{figure}
	\center
	\includegraphics[width=0.5 \linewidth]{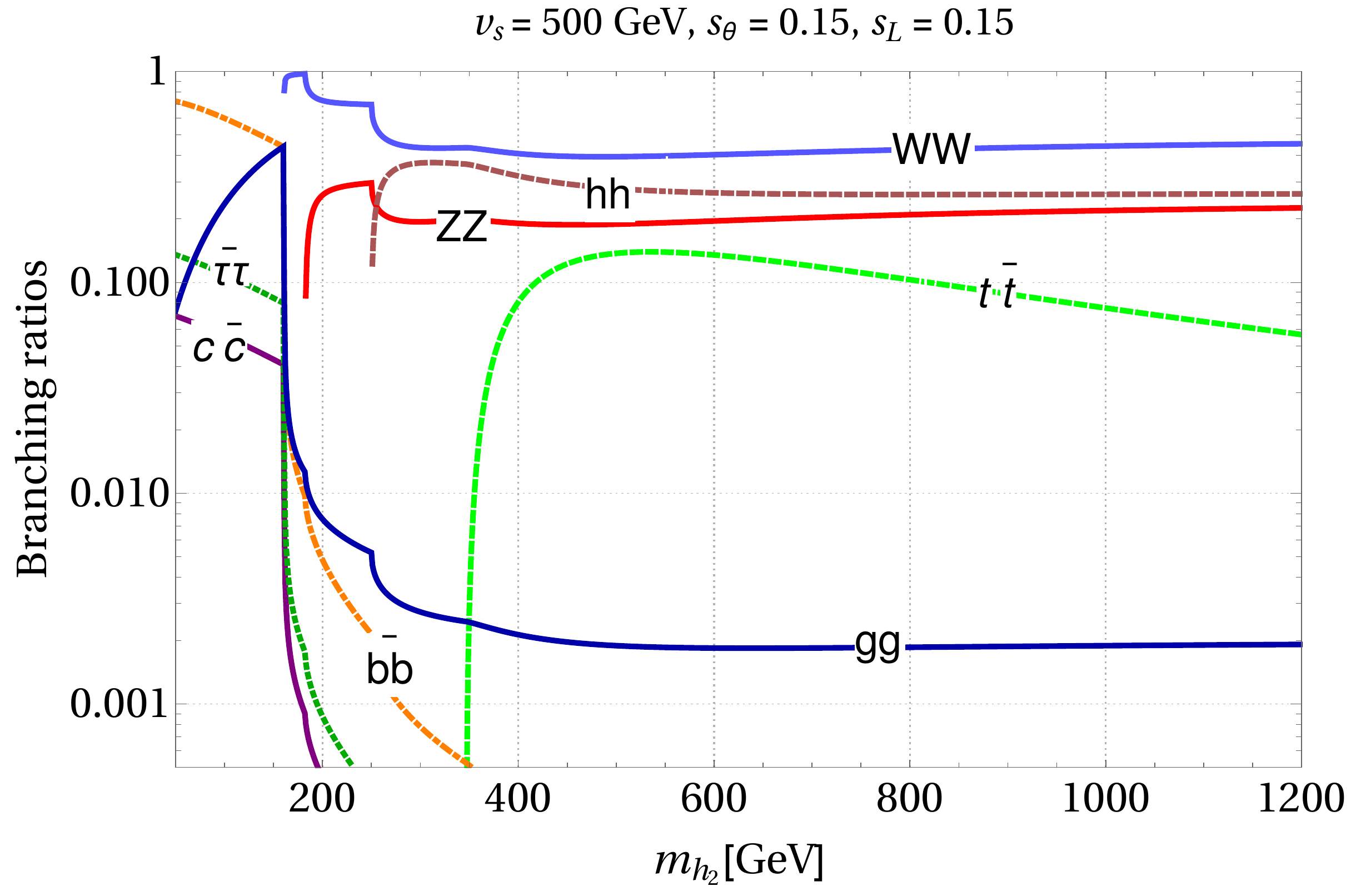}
	\caption{\small Branching fractions of the new scalar as a function of $m_{h_2}$ for $s_{\theta}=s_{L}=0.15$ and $v_S=500$ GeV.}
	\label{BR}
\end{figure}

Based on the decay modes of $h_{2}$, different topologies are 
expected to be produced.
The $h_{2}$ decay mechanism is almost similar to the Higgs boson, dominated by the $b\bar{b}$ pair and 
two gluons in low mass region while the decay channels $WW$, $ZZ$, $t\bar{t}$, and $hh$ are dominant at large
mass region. 
For more illustration,  the  branching fractions of the $h_{2}$ decays into the SM particles
are shown in Fig.\ref{BR} for $v_{S} = 500$ GeV and $s_\theta = s_{L} = 0.15$.  With increasing $m_{h_2}$, decay modes of $WW$, $ZZ$, and $hh$ are dominant with respect to $t\bar{t}$.
From Fig.\ref{BR}, one can see that at $m_{h_2} \geq200\,$GeV, decay channels $h_2\rightarrow gg/\gamma\gamma/f\bar f$,  
where $f$  denotes all fermion except top quark, are negligible.

In this section, we derive potential bounds on the free parameters of the simplified model in $e^- e^+$ colliders through $h_{2}+\gamma$ 
final state and compare 
them with the ones obtained from Higgs and precision electroweak data measured at the LHC \cite{Dolan:2016eki}. 
Given the interested mass range of $h_{2}$, some of the appropriate,
dominant, and clean decay modes are listed in Table \ref{decays}
as a reference for discussion of the experimental signatures. 

\begin{table}[]
\centering
\caption{ The $h_{2}$ to $WW$, $hh$, $ZZ$, and $t\bar{t}$ decay channels as well as the branching fractions of some of the decay modes.}
\label{decays}
\begin{tabular}{c|c|c|c} \hline\hline
  $ZZ$  mode [$Br ~  \%$]      &  $t\bar{t}$ mode [$Br ~ \%$] &   $WW$ mode [$Br ~ \%$] &  $hh$ mode [$Br ~ \%$]                     \\  \hline
  $(ll)(l'l')[1\%]$                    &   $(l\nu_{l} b)(q\bar{q}'b)[44\%]$     & $(l\nu_{l})(q\bar{q}')[44\%]$  &   $(b\bar{b})(b\bar{b})[34\%]$                    \\
 $(ll)(q\bar{q})[14\%] $            &    $(l\nu_{l} b)(l'\nu_{l'} b)[10\%]$     & $(l\nu_{l})(l'\nu_{l'} )[10\%]$   &   $(b\bar{b})(ll)[7\%]$                            \\
 $(ll)(\nu_{l'}\nu_{l'})[4\%]$      & $(q\bar{q}'b)(q\bar{q}'b)[46\%]$    &  $(q\bar{q}')(q\bar{q}')[46\%]$   &   $(ll)(ll)[0.4\%]$                                    \\ 
   $(q\bar{q})(q'\bar{q'})[49\%]$ &  $.....$                                   &  $..... $     &   $(b\bar{b})(\gamma \gamma)[0.3\%]$                                        \\  
      $(q\bar{q})(\nu\bar{\nu})[28\%]$ &  $.....$                                   &  $..... $    &   $(ll)(\gamma \gamma)[0.03\%]$                                          \\  
   $(\nu\bar{\nu})(\nu\bar{\nu}) [4\%]$ &  $.....$                                   &  $..... $   &    $(WW)(WW)[5\%]$ \\ 
                                        \hline\hline
\end{tabular}
\end{table}

In order to make an estimation of the potential of $e^- e^+  \rightarrow h_{2}+\gamma$ process to 
probe the parameter space of the model, we consider the $h_{2}$ decay into $hh$, followed by $h\rightarrow b\bar{b}$, {\it i.e.}
$e^- e^+  \rightarrow h_{2}\gamma \rightarrow hh\gamma \rightarrow b \bar{b} b \bar{b} \gamma$. These choices are made
due to the large branching fractions of $h_{2} \rightarrow hh$ and $h\rightarrow b\bar{b}$. 
The cross section of signal including the whole decay chain is computed as 
$\sigma(e^+ e^-  \rightarrow h_{2}\gamma) \times Br(h_2\rightarrow h h)\times Br(h\rightarrow b\bar{b})^2$. 
For $m_h=125\gev$, the Higgs boson branching ratio decaying into $b \bar{b}$ is $0.584$\cite{Olive:2016xmw}.
We note that the initial state radiation (ISR)  affects the signal and background cross sections. In calculating the cross sections, 
the Jadach-Ward-Was formalism \cite{ward}
is used to consider the ISR effects.  
%It is found that switching on the ISR changes the signal efficiency by $3\%$.

The main background processes to this final state are:
\begin{eqnarray}
&&e^{+}e^{-} \rightarrow  4b+\gamma ,\nonumber \\
&&e^{+}e^{-} \rightarrow  2b+2j+\gamma~(j=u,d,s,c,g), \text{ two jets $(j)$ are misidentified as b-jets},\nonumber \\
&&e^{+}e^{-} \rightarrow 4j+\gamma~(j=u,d,s,c,g),\text{four jets ($j$) are misidentified as b-jets},  
\end{eqnarray}
where all processes with off-shell $\gamma, Z, W^{\pm}$and gluons are included.

The background rates are computed using {\sc CompHEP} \cite{comphep} and  {\sc MadGraph5-aMC@NLO} packages \cite{mg}.
The efficiency of b-tagging for a jet originating from the hadronization of a bottom quark is taken $70\%$, and
misidentification rates of $10\%$ and $1.5\%$ are assumed for charm quark and light-flavor jets \cite{Arominski:2018uuz}. 
To select signal and background events and achieve a good sensitivity, the following requirements
are applied: energy ($E$) and
pseudorapidity ($\eta = -\ln\tan(\frac{\theta}{2})$) of b-jets are required to be larger than 20 GeV and $|\eta_{\rm b-jet}| < 2.5$, respectively.
Since the photon is expected to carry a large amount of energy due to its recoil against the heavy scalar $h_{2}$,
its energy is required to be greater than 300 GeV and $|\eta_{\gamma}| < 2.5$. 
The signal and background efficiencies after these cuts are found to be $20.02\%$ and $0.67\%$, respectively.
As in the signal events, the $b\bar{b}$  pairs come from the Higgs boson decays,
it is required $100~ \text{GeV} \leq m_{b\bar{b}} \leq 150$ GeV. This requirement 
has a strong power to suppress the background processes where the final state 
jets are not originating from Higgs bosons decays. For instance, it  provides a 
rejection rate  at the order of $\lesssim 10^{-5}$ for the major background processes, {\it i.e.}
 $e^{-}+e^{+} \rightarrow 4b+\gamma$.

It should be mentioned that there are background processes in which an additional jet is misidentified as a photon.
Such a signature may occur when neutral pions with large boost appear from 
jet fragmentation and decay to two photons. 
The showers from two photons can overlap in the electromagnetic calorimeter (ECAL)
and will  be observed as a single photon.
The probability for a jet to be misidentified as a photon is dependent on the  photon energy
and is of the order of $10^{-5,-4}$ \cite{Ellis:2012zp}
for an energetic fake photon. Therefore,
requiring a photon with $E \geq 300$ GeV suppresses the fake background contribution
to a negligible level. A realistic detector simulation is necessary to estimate the
fake background contribution.

\begin{figure}
	\center
		\includegraphics[width=0.4 \linewidth]{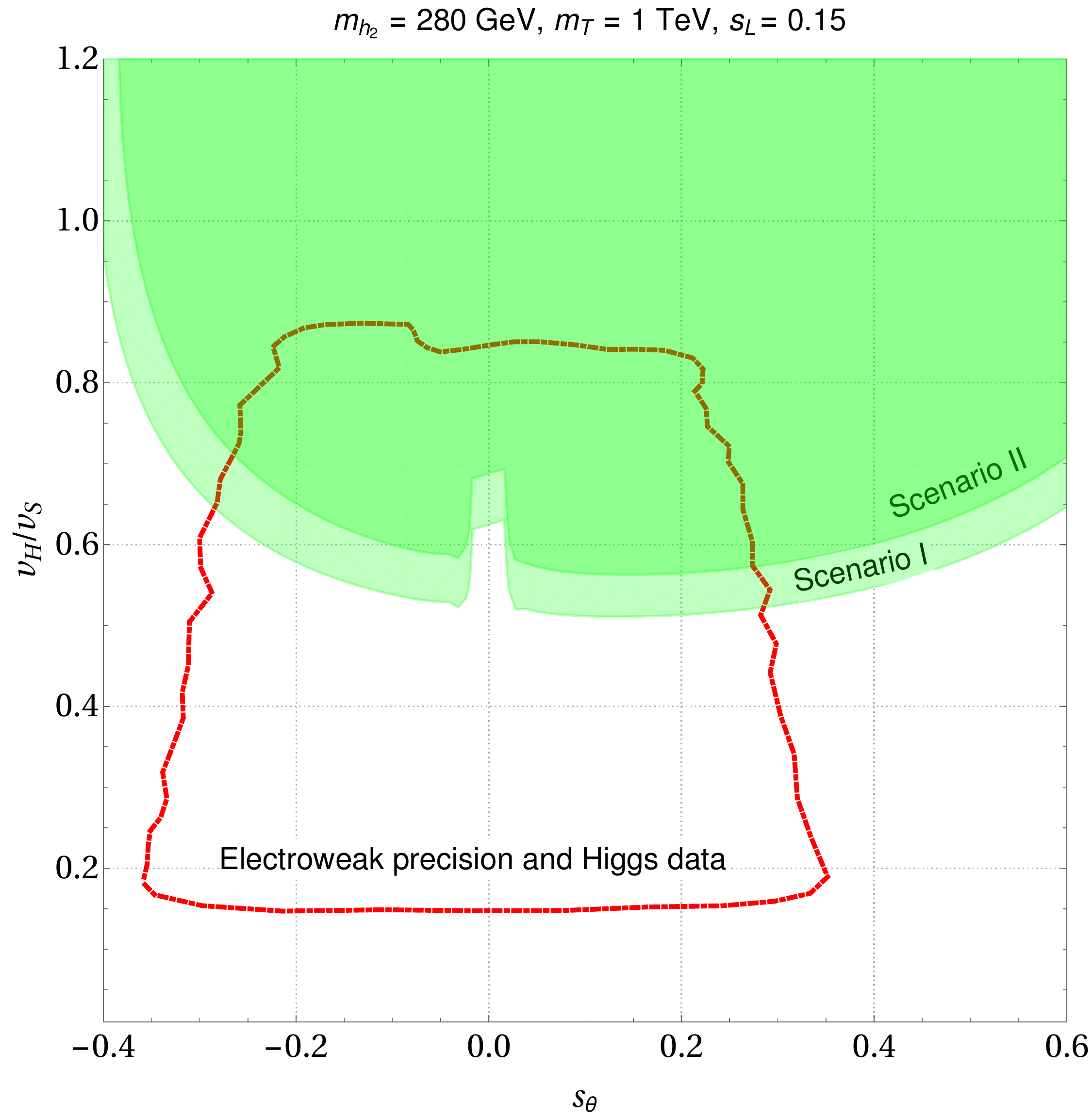}
			\includegraphics[width=0.4 \linewidth]{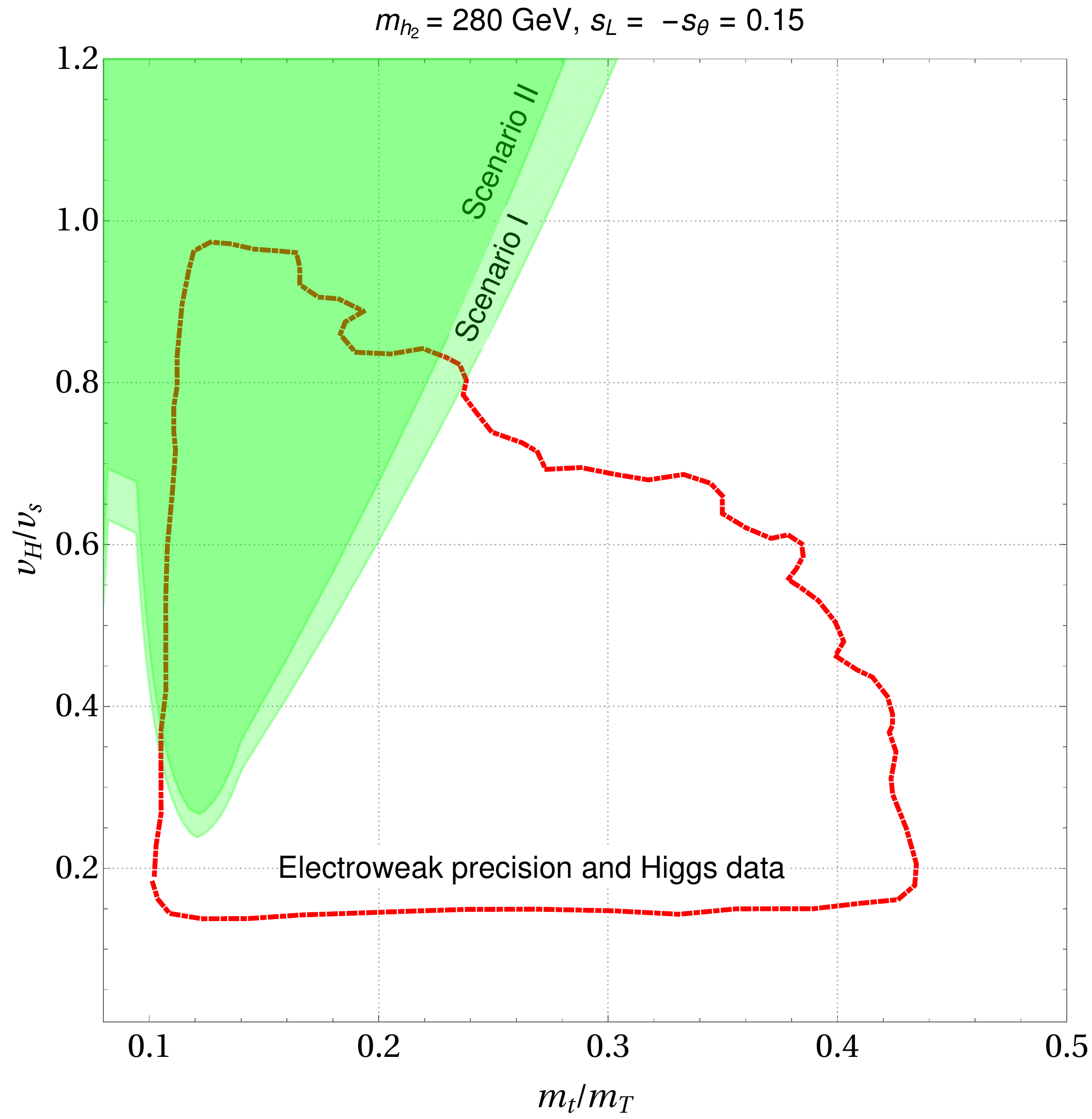}
				\caption{\small The green areas are the $95 \%\,$CL  excluded  regions of the parameter space
				 in the planes of ($v_H/v_S$,$s_\theta$) (left)
	and ($v_{H}/v_{S}$,$m_t/m_T$) (right) for $\sqrt{s}=3\,$ TeV with $\mathcal{L}=3 \inab$. The results are presented assuming that 
	$m_{T} = 1$ TeV and $m_{h_2} = 280$ GeV.  
	 The dot-dashed red curves represent contours of favoured region at $95 \%\,$CL 
	  obtained from the precision electroweak and Higgs data from LHC \cite{Dolan:2016eki}.}
	\label{res}
\end{figure}

In Fig.\ref{res}, the $95 \%\,$CL excluded regions for  $v_H/v_S$ versus $s_\theta$ and $v_H/v_S$  versus $m_t/m_T$
for the center-of-mass energy of 3 TeV with 
an integrated luminosity of 3 ab$^{-1}$ are presented. The excluded regions are depicted for two scenarios 
of background contributions: scenario I where no uncertainty is considered on the
number of expected background and scenario II where an uncertainty of $50\%$ is taken 
on the number of expected backgrounds.
The dot-dashed red curves in Figs.\ref{res} show the contours of favoured region at $95 \%\,$CL 
extracted from the precision electroweak data and Higgs boson measurements at the LHC \cite{Dolan:2016eki}.
As it can be seen, with the proposed selection in this work, for any value of the sine of mixing angle $s_{\theta} \in [0.05,0.3]$,
any value of  $v_{H}/v_{S}$ above $0.55$, corresponding to $v_{S} \lesssim 450$ GeV, can be excluded. This would be 
a considerable improvement with respect to current bounds from electroweak precision tests and LHC Higgs data. 
The  $95 \%\,$CL excluded region for $v_{H}/v_{S}$ versus $m_{t}/m_{T}$ 
indicates that a part of allowed region from electroweak precision tests and Higgs data corresponding to
large mass of VLQ is accessible via the $h_{2}\gamma$ channel. For example, any value of  $v_{S} \lesssim 1  $ TeV
can be excluded for $m_{T} = 1.4$ TeV.

\begin{figure}
 	\center
     \includegraphics[width=0.4 \linewidth]{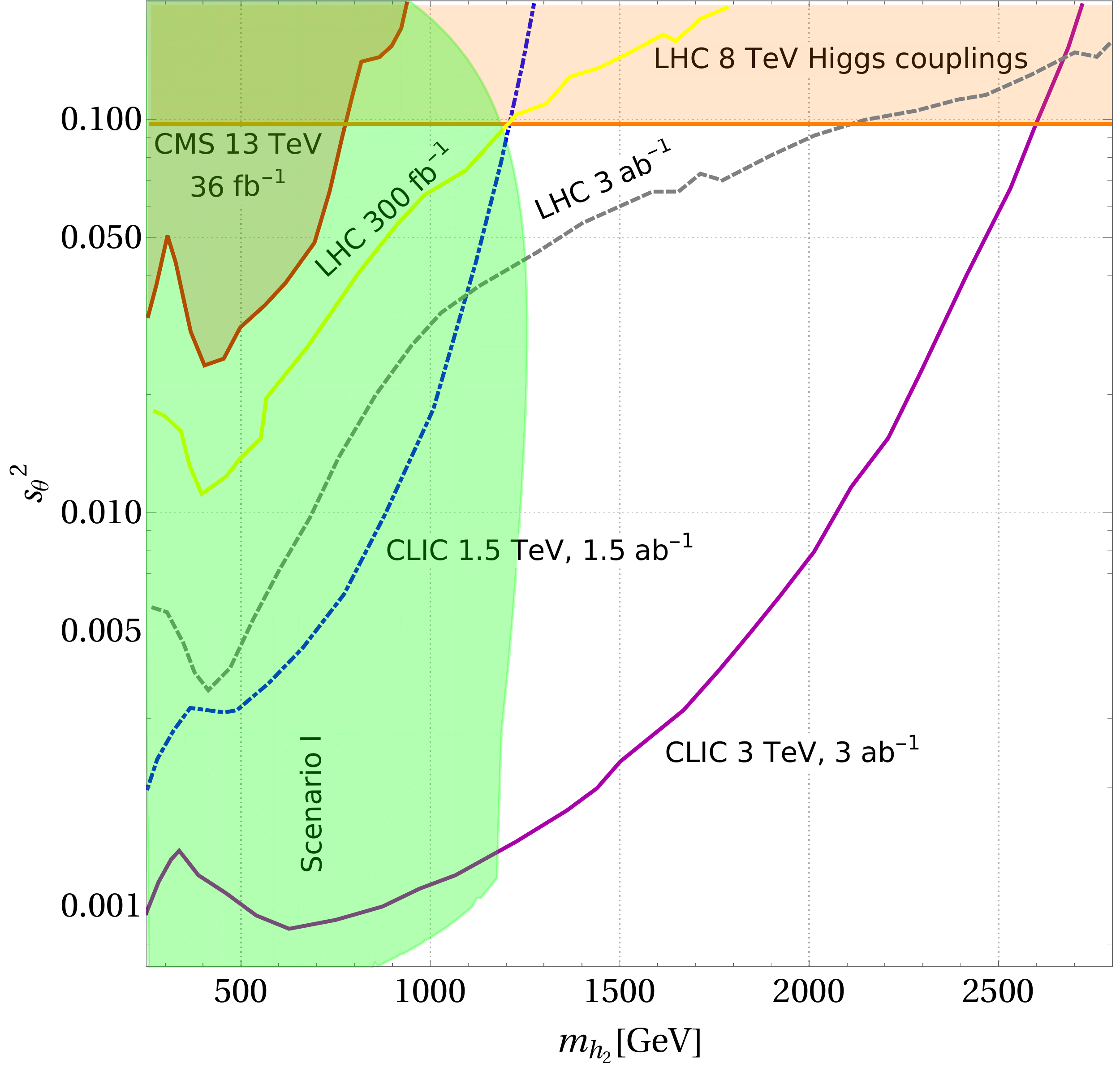}
 	\includegraphics[width=0.4 \linewidth]{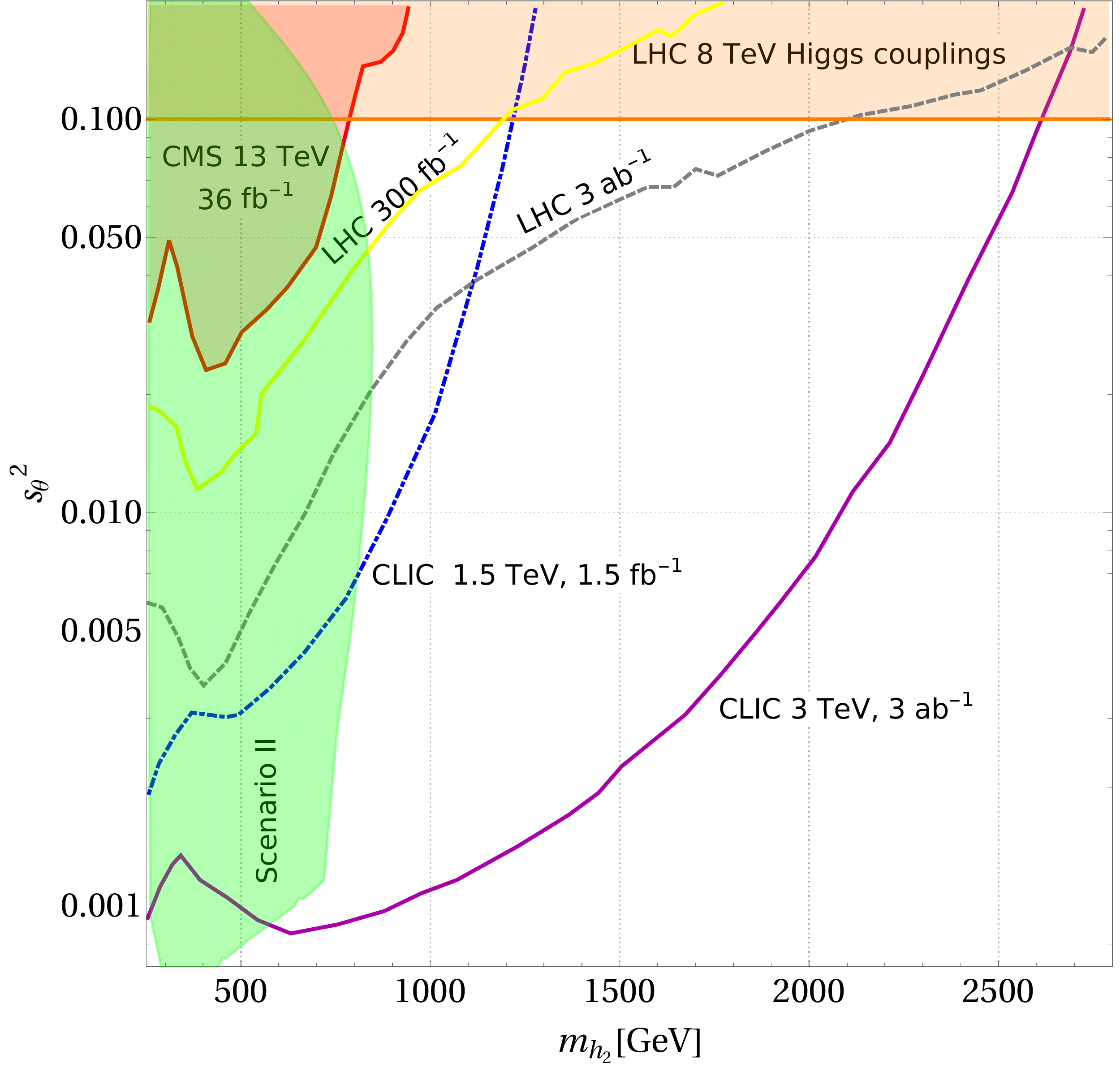}
 	\caption{\small The green areas are the $95 \%\,$CL  excluded regions of the parameter space in the plane of ($s_\theta^2,m_{h_{2}}$)
	 with no (left) and with $50\%$ (right)  uncertainty on the expected number of background events.
	 These regions are derived from $e^- e^+ \rightarrow h_{2}\gamma$ with subsequent $h_{2} $
	  decay into di-Higgs and Higgs decays into $b\bar{b}$.  The results are shown under assumption of  $m_{T} = 1$ TeV, 
	  $s_L=0.15$ and $v_S=350 \gev$. The red and orange shaded regions obtained from LHC direct search with
	  the scalar decays into di-boson and from indirect searches. The constraints from CLIC at $\sqrt{s} = 1.5 \tev$ and $1.5 \inab$ 
	   (dot-dashed blue) and $\sqrt{s} = 3 \tev$ and $3 \inab$, obtained from the single scalar channel, are depicted with blue and purple curves. 
	   The constraints from the HL-LHC with a luminosity of $300 \infb$ (yellow) and  $3\inab$ (dashed gray) are presented for comparison as well  \cite{Buttazzo:2018qqp}.
 		}
 	\label{res22}
 \end{figure}

In  Fig. \ref{res22}, the  excluded regions of the parameter space
 in the plane of ($s_\theta^2,m_{h_{2}}$) at
 $95 \%\,$CL are shown.
 The results are presented for two scenarios of the expected background and are
 compared with those derived from 36 fb$^{-1}$ of LHC data at 13 TeV,  $e^{-}e^{+} \rightarrow h_{2}+\nu\bar{\nu}$ process \cite{Buttazzo:2018qqp}
 at CLIC, and the high luminosity LHC (HL-LHC) \cite{Buttazzo:2018qqp}.  As it can be seen, using the
  $e^{-}e^{+} \rightarrow h_{2}+\nu\bar{\nu}$ process at the center-of-mass energy of 3 TeV with 
 an integrated luminosity of 3 ab$^{-1}$, any value of $s^{2}_{\theta}$ above $\sim 0.0015$ for $m_{h_{2}} \sim 300$ GeV 
 could be excluded.  The HL-LHC would be able to exclude $s^{2}_{\theta} \gsim 0.005$ for $m_{h_{2}} \sim 300$ GeV
 using the scalar decay into di-boson \cite{Buttazzo:2018qqp}. The associated production of $h_{2}$ 
 with a photon is sensitive to a mass region of $m_{h_{2}} \sim 1200$ GeV for the values of $s^{2}_{\theta} \lesssim 0.15$.
 Considering no uncertainty would provide access to a mass region of  $m_{h_{2}} \lesssim 1200$ GeV and taking into account 
 an uncertainty of  $50\%$ on the expected background  would  reduce the sensitivity to mass, with almost 800 GeV
 for $s^{2}_{\theta} \lesssim 0.1$.
As we can see, the $h_{2}\gamma$ process would be able to scan a remarkable region of $s^{2}_{\theta} \lesssim 0.02$ with
$m_{h_{2}} \lesssim 800$ GeV where the HL-LHC is not sensitive to. 
From Fig. \ref{res22}, we  also see that the excluded region derived from 
$h_{2}\gamma$ plays a complementary role to single scalar production at CLIC \cite{Buttazzo:2018qqp}. 
In particular, the $h_{2}\gamma$ process could probe the model at  a mass region of $m_{h_{2}} \lesssim 1100$ GeV, 
with $s^{2}_{\theta} \lesssim 0.001$ which is not accessible
by the single scalar production.

The presented results here are at the center-of-mass energy of 3 TeV and
are based on an integrated luminosity of 3$\inab$ which is expected to be attainable
by CLIC according to Ref.\cite{Abramowicz:2016zbo}. However, it should be noted that 
different center-of-mass energies are sensitive to a particular part of the model parameters.
For example,  as the left plot of Fig.\ref{sigma-mh2}  shows, the total cross section of $h_{2}\gamma$ process 
is maximum at  $\sqrt{s} = 1$ TeV for  $m_T=v_S=500\,$ GeV  while with $m_T= 1 $ TeV, $v_S=350\,$ GeV,
 collisions at $\sqrt{s} = 3$ TeV provides
the largest cross section.  Therefore, performing the study at different center-of-mass energies
and combination would allow to extend the sensitivity to larger part of the parameter space.

In this investigation, we only concentrated on di-Higgs decay mode of $h_2$, followed by $h\rightarrow b \bar{b}$, and cuts on 
few kinematic variables are considered to reduce the backgrounds. However, there are rooms for improving
 the sensitivity which could be achieved by {\it (i)} considering other decay modes
 of the Higgs boson as mentioned in the last column of Table \ref{decays}; {\it (ii)} using the other decay channels of 
 $h_{2}$ like $h_{2} \rightarrow WW, ZZ, t\bar{t}$ and combination of all channels together;
 {\it (iii) }
  using various kinematic variables to suppress the contributions and also 
 exploiting sophisticated methods such as multivariate techniques to distinguish signal events
 from background events.

%
%%%%%%%%%%%%%%%%%%%%%%%%%%%%%%%%%%%%%%%%%%    Summary and remarks    %%%%%%%%%%%%%%%%%%%%%%%%%%%%%%%%%%%%%%%%%%%%%%%%%
\section{ Summary}\label{sec:conclu}
%%%%%%%%%%%%%%%%%%%%%%%%%%%%%%%%%%%%%%%%%%%%%%%%%%%%%%%%%%%%%%%%%%%%%%%%%%%%%%%%%%%%%%%%%%%%%%%%%%%%%
%
To summarise, we performed an analytic calculation for the production cross section 
of a neutral singlet scalar heavier than the observed SM Higgs boson
in association with a photon ($h_{2}\gamma$) in the context of a simplified SM extension model, at a lepton
collider. In this model the scalar mixes with the SM Higgs boson and  can couple to a pair of
SM massless bosons, {\it i.e.} photon and gluon, at loop level 
via a vector like fermion which mixes with the top quark. 
Such a model provides the possibility to  stabilise the electroweak vacuum\cite{Xiao:2014kba}.

We presented the size of the $h_{2}\gamma$ cross section for various center-of-mass energies
which would be reached at future lepton colliders. The cross section is calculated in 
terms of the model parameters and in general it was found to be small but the signal signatures are rather clean
and could be used to span the parameter space of the model. To examine the potential of the process
to explore the model, we concentrated on the decay of $h_2$ to $hh$ followed by the Higgs bosons decays into $b\bar{b}$ pairs.
Using a set of selection cuts for the signal events which efficiently suppress the background contributions,
the exclusion regions at $95\%$ CL are derived. This was done as an example at the center-of-mass energy of $\sqrt{s} = 3$ TeV with 
an integrated luminosity of 3 ab$^{-1}$. 
Overall, the results indicate that the associated production of the scalar $h_2$
with a photon is sensitive to a part of the parameter space with $m_{h_2} \lesssim 1.2$ TeV and low values of the
mixing angle which is out of access at the HL-LHC.  A comparison of the results with the single scalar production
at CLIC suggests that the $h_{2}\gamma$
process has the ability to  complement the single scalar channel at very low values of the mixing angle.

%%%%%%%%%%%%%%%%%%%%%%%%%

\vspace{1cm}
%%%%%%%%%%%%%%%%%%%%%%%%    Acknowledgments    %%%%%%%%%%%%%%%%%%%%%%%%%%%%%%
%
{\bf Acknowledgments:}
M. Mohammadi Najafabadi is grateful to INSF for the support. 
S. Tizchang is grateful to Hiren Patel and  Rodolfo Capdevilla for valuable help in calculations with packages in Mathematica,
 and F. Elahi for fruitful discussions. 
%%%%%%%%%%%%%%%%%%%%%%%%%%%%%%%%%%%%%%%%%%%%%%%%%%%%%%%%%%%%%%%%%%%%%%%%%%%%%%%%%%%%%%%%%%%%%%%%%%%%%
%

%%%%%%%%%%%%%%%%%%%%%%%%% %%%%%%%%%%%%%%%%%%%%%%       Appendix       %%%%%%%%%%%%%%%%%%%%%%%%%%%%%%%%%%%%%%%%%%%%%%%%%%%%%%%%%%%%%%%%%
\begin{appendix}
	\section{Appendix}\label{app}
%%%%%%%%%%%%%%%%%%%%%%%%%%%%%%%%%%%%%%%%%%%%%%%%%%%%%%%%%%%%%%%%%%%%%%%%%%%%%%%%%%%%%%%%%%%%%%%%%%%%%%
%
The $F$ functions appeared in  Eq.\ref{formfi} are defined as:
\begin{eqnarray}
F^a&=& 4\Big[ \mathbf{C}_{22}^a+\mathbf{C}_{12}^a+\mathbf{C}_{2}^a+\frac{\mathbf{C}_0^a}{4}\Big],\nonumber\\
F^{n}&=& 2\, (m_t+ m_T)
(\mathbf{C}_{22}^n+
\mathbf{C}_{12}^n)+m_T\,
\mathbf{C}_0^n+(m_t+3\, m_T)
\mathbf{C}_2^n-(m_t-
m_T)~
\mathbf{C}_1^n,~~~~~~\nonumber\\
F^{l}&=&F^{n}\Big[m_t\leftrightarrow m_T,n\rightarrow l\Big],
\end{eqnarray}     
where $a=f,t,T$ and $\mathbf{C}_{ij}$s  are the  common scalar two and three point Passarino-Veltman functions which are defined as:        
\begin{eqnarray}
\mathbf{C}_{ij}^f &\equiv& \mathbf{C}_{ij}(s,0,m_{h_2}^2,m_f^2,m_f^2,m_f^2),\nonumber\\
\mathbf{C}_{ij}^t &\equiv& \mathbf{C}_{ij}(s,0,m_{h_2}^2,m_t^2,m_t^2,m_t^2),
\nonumber\\
\mathbf{C}_{ij}^T &\equiv& \mathbf{C}_{ij}(s,0,m_{h_2}^2,m_T^2,m_T^2,m_T^2),
\nonumber\\
\mathbf{C}_{ij}^{n} &\equiv& \mathbf{C}_{ij}(s,0,m_{h_2}^2,m_t^2,m_T^2,m_T^2),
\nonumber\\
\mathbf{C}_{ij}^{l} &\equiv& \mathbf{C}_{ij}(s,0,m_{h_2}^2,m_T^2,m_t^2,m_t^2).
\end{eqnarray}  
In Eq.\ref{formfi}, $ F^{\gamma/Z,W}$ are the combination of  Passarino-Veltman  scalar functions:
\begin{eqnarray}
F^{\gamma,W} &=&
4\left( \frac{m_{h_2}^2}{M_W^2} +6 \right)  (  \mathbf{C}_{12} + \mathbf{C}_{22} + \mathbf{C}_2) + 16  \mathbf{C}_0, \nonumber \\
F^{Z,W} &=&2 \left[ \frac{m_{h_2}^2}{M_W^2} (1-2 c_w^2)   + 2 ( 1-6 c_w^2 ) \right]( \mathbf{C}_{22}+ \mathbf{C}_{12}+ \mathbf{C}_{2} )+4 (1-4 c_w^2) \mathbf{C}_{0},
\end{eqnarray}
with
\begin{eqnarray}
\mathbf{C}_{ij} &\equiv& \mathbf{C}_{ij}(s,0,m_{h_2}^2,M_W^2,M_W^2M_W^2).
\end{eqnarray}

The $C^{\pm{\rm Wbox}}_{1,2}$ and $C^{\pm{\rm Zbox}}_{1,2}$ in Eq.\ref{formfactor} have the following forms:

\begin{eqnarray}
C^{+{\rm Wbox}}_{1}&=& -\frac{e^4 M_W~s_{\theta}}{s_w^3}\Big[\mathbf{D}_1^a+\mathbf{D}_1^b+\mathbf{D}_{13}^b-\mathbf{D}_{13}^a-\mathbf{D}_{33}^a\Big],\nonumber\\
C^{+{\rm Wbox}}_{2}&=& -\frac{e^4 M_W~s_{\theta}}{s_w^3}\Big[\mathbf{D}_2^a+\mathbf{D}_{23}^a+\mathbf{D}_2^b-\mathbf{D}_{23}^b-\mathbf{D}_{33}^b\Big],\nonumber\\
C^{-{\rm Wbox}}_{1,2}&=&0,\nonumber\\
C^{\pm{\rm Zbox}}_{1}&=&- \frac{2~ e^4 M_W~s_{\theta}~g_e^{\mp 2}}{ s_wc_w^2}\Big[\mathbf{D}_{13}^c+\mathbf{D}_{33}^c\Big],\nonumber\\
C^{\pm{\rm Zbox}}_{2}&=&\frac{2~ e^4 M_W~s_{\theta}~g_e^{\mp 2}}{ s_wc_w^2}\Big[\mathbf{D}_2^c+\mathbf{D}_{12}^c+\mathbf{D}_{23}^c\Big],
\end{eqnarray}
where $g_e^-\equiv I^3_e/(s_wc_w)-Q_e s_w/c_w$, and $ g_e^+\equiv-Q_e s_w/c_w$. Four point Passarino-Veltman functions $\mathbf{D}_{ij}$ are defined as follows:
\begin{eqnarray}
\mathbf{D}_{ij}^{a}&\equiv& \mathbf{D}_{ij}(0,s,m_{h_2}^2,u,0,0,0,M_W^2,M_W^2,M_W^2),\nonumber\\
\mathbf{D}_{ij}^{b}&\equiv& \mathbf{D}_{ij}(0,s,0,t,0,m_{h_2}^2,0,M_W^2,M_W^2,M_W^2), \nonumber\\
\mathbf{D}_{ij}^{c}&\equiv&\mathbf{D}_{ij}(0,u,m_{h_2}^2,t,0,0,0,0,M_Z^2,M_Z^2). \label{Coll:div:potential}
\end{eqnarray}
\end{appendix}

\end{document}